# A Robust Pipeline for Differentially Private Federated Learning on Imbalanced Clinical Data using SMOTETomek and FedProx

**Rodrigo Tertulino** [ **IFRN** | ***rodrigo.tertulino@ifrn.edu.br*** ]

*Federal Institute of Education, Science, and Technology of Rio Grande do Norte (IFRN), Software Engineering and Automation Research Laboratory - LaPEA, 59628-330, Mossoró-RN, Brazil.*

**Abstract** Federated Learning (FL) offers a groundbreaking approach to collaborative health research, enabling model training on decentralized data while safeguarding patient privacy. FL offers formal security guarantees when combined with Differential Privacy (DP). The integration of these technologies, however, introduces a significant trade-off between privacy and clinical utility, a challenge further complicated by the severe class imbalance often present in medical datasets. The research presented herein addresses these interconnected issues through a systematic, multi-stage analysis. An FL framework was implemented for cardiovascular risk prediction, where initial experiments showed that standard methods struggled with imbalanced data, resulting in a Recall of zero. To overcome such a limitation, we first integrated the hybrid **Synthetic Minority Over-sampling Technique with Tomek Links (SMOTETomek)** at the client level, successfully developing a clinically useful model. Subsequently, the framework was optimized for non-IID data using a tuned **FedProx** algorithm. Our final results reveal a clear, non-linear trade-off between the privacy budget ($\epsilon$) and model utility. An optimal operational region was identified where formal differential privacy guarantees with a single-digit privacy budget ($\epsilon \approx 9.0$) can be achieved while maintaining high clinical utility (Recall $\approx 77\%$) and discriminative power (ROC-AUC $\approx 0.82$). Although single-digit $\epsilon$ values are more relaxed than the tight budgets studied in theoretical DP literature, they represent a pragmatic operating point commonly reported in applied healthcare FL research. Ultimately, our study provides a practical methodological blueprint for creating effective, secure, and accurate diagnostic tools that can be applied to real-world, heterogeneous healthcare data.

## 1 Introduction

Cardiovascular diseases (CVDs) remain the leading cause of global mortality, responsible for an estimated 20.5 million deaths in 2021 alone Mensah *et al*. [2023]. The scale of this health crisis underscores the urgent need for effective early-detection and risk-prediction tools. Machine learning (ML) has emerged as a powerful approach to meet this need, demonstrating its potential to predict cardiovascular events by identifying complex patterns across numerous clinical and lifestyle risk factors Khimani *et al*. [2024]; Paul *et al*. [2021].

However, the efficacy of these models depends on access to large-scale, diverse datasets, a requirement that is often hindered by the sensitive nature of patient health information. Strict data protection regulations, such as the General Data Protection Regulation (GDPR) Mittal *et al*. [2024], and the Health Insurance Portability and Accountability Act (HIPAA) Jensen *et al*. [2007], lead to the formation of "data silos". Consequently, valuable data remains isolated within individual institutions, severely limiting the potential for collaborative research Lopez *et al*. [2023].

Federated Learning (FL) has emerged as a powerful solution to the aforementioned challenge of data silos. Therefore, FL is a distributed learning paradigm where multiple clients (e.g., hospitals) collaboratively train a global model under the coordination of a central server, without ever sharing their raw data Abhishek *et al*. [2023]. Each client trains the model on its local data and only communicates the resulting parameter updates (e.g., weights and gradients) to the server for aggregation and fusion. Moreover, the privacy-preserving approach enables institutions to build more robust and generalizable models by leveraging a collective dataset, a feat that is unachievable in isolation Gupta *et al*. [2023].

While FL provides a strong privacy baseline, the model updates are not immune to sophisticated attacks that could infer sensitive information about the training data Nguyen *et al*. [2025]. Therefore, Differential Privacy (DP) is often integrated into the FL pipeline to guarantee formal, mathematical privacy. Differential Privacy ensures that the contribution of any single individual in the dataset is statistically obscured, typically by adding calibrated noise to the model updates before they are sent to the server Khimani *et al*. [2024]. The injection of noise for privacy, however, inherently introduces a trade-off: stronger privacy guarantees often come at the cost of reduced model accuracy and utility Thumula *et al*. [2025]. As noted by Dubey *et al*. [2025], implementing DP can lead to a measurable drop in model performance, creating a critical dilemma in clinical applications where high accuracy is paramount.

The present work goes beyond simply recognizing the privacy-utility trade-off; it documents the methodological steps required to achieve and optimize it in a realistic and challenging healthcare setting. While many studies apply FL to healthcare, they ignore the cumulative effects of class imbalance and statistical heterogeneity. Consequently, the primary goal is to present a robust, multi-stage pipeline that enables a clinically viable model and then systematically analyze its performance under different levels of DP. We place primary emphasis on the **Recall** metric, as failing to identify a high-risk patient (a false negative) has the most severe clinical consequences. However, to ensure a comprehensive evaluation, we also analyze **ROC-AUC**, **Precision**, and **F1-score**, providing a holistic view of the model's discriminative power and the inherent trade-offs in privacy-preserving screening. The contributions of our work are threefold:

- **Demonstration of a Critical Failure Case:** We first establish that a standard FL with DP application on a severely imbalanced clinical dataset fails to produce a useful model, yielding misleadingly high accuracy but zero clinical Recall. This serves as a critical baseline and a cautionary finding for practitioners.
- **A Validated Pipeline for Achieving Clinical Utility:** To overcome such a failure, we propose and validate a robust methodological pipeline. It involves integrating locally applied client-side SMOTETomek to address class imbalance and employing the FedProx algorithm to mitigate the effects of non-IID data, resulting in a model with significant predictive power.
- **Robust Statistical Analysis of the Privacy-Utility Frontier:** We present a granular analysis of the privacy-utility trade-off, validated through multiple independent experimental runs to account for the stochasticity of DP. We identify an optimal operational region where the model achieves robust discriminative power (Mean ROC-AUC $\approx 0.82$) and high sensitivity (Recall $\approx 77\%$) under formal DP guarantees with a single-digit privacy budget ($\epsilon \approx 9.0$), providing a statistically significant blueprint for deploying secure FL systems in healthcare.

The paper is structured as follows: **Section 2** reviews the relevant literature, analytically distinguishing our contribution from existing frameworks in healthcare FL. **Section 3** outlines the problem formulation, focusing on the challenges of statistical heterogeneity and privacy leakage. **Section 4** details the multi-stage experimental methodology, specifying the local data balancing strategy, the FedProx aggregation, and the differential privacy accounting. **Section 5** presents the empirical results, validated through robust statistical analysis, and discusses the privacy-utility trade-off and the model's discriminative capability. Finally, **Section 6** concludes the paper by summarizing our findings and outlining future research directions. An **appendix** is also provided, containing reproducible code snippets of the core pipeline functions.

# 2 Related Work

The literature on FL has grown exponentially, with significant efforts focused on its application in sensitive domains, such as healthcare, on the persistent challenges of heterogeneity, and on the crucial role of Privacy-Enhancing Technologies (PETs). A summary of the articles is presented in **Table 1**.

## 2.1 Federated Learning in Healthcare

The application of FL in medicine has moved from theoretical concepts to practical benchmarks and specific use cases. Moreover, a significant advancement for applying FL to healthcare is the introduction of FedCVD by Zhang *et al*. [2024], the first real-world FL benchmark focused on CVD data, providing a standardized basis for evaluating new algorithms. The scope of FL in healthcare is broad, with many studies proposing frameworks for specific conditions. For example, Gaber *et al*. [2024] introduced a distinct framework, also named FedCVD, a scalable FL model using logistic regression and SVM for CVD prediction, demonstrating competitive performance against centralized models while ensuring data privacy.

Furthermore, Ahmed *et al*. [2025] proposed a dynamic, scoring-based client selection mechanism for a federated diabetes diagnosis system, emphasizing the importance of intelligent participant selection to enhance model performance and fairness. Comprehensive surveys, such as the one by Hudaib *et al*. [2025], review the implementation of deep learning models within FL for various healthcare tasks, including an emphasis on medical imaging applications. These works collectively underscore the field's rapid maturation and its potential to revolutionize medical data analytics while navigating the complexities of data privacy and heterogeneity.

## 2.2 Key Challenges in Federated Learning

The practical deployment of FL is complicated by several inherent challenges, primarily categorized as system and statistical heterogeneity. System heterogeneity refers to the diversity in clients' computational power, network bandwidth, and availability. The bottleneck is the "straggler" problem, where slower clients delay synchronous FL protocols. Hence, several solutions have been proposed to mitigate the issue. For instance, Nishio and Yonetani [2019] developed FedCS, a client selection protocol that prioritizes clients with sufficient resources to complete training within a given deadline. Another approach is FLANP, introduced by Reisizadeh *et al*. [2022], which is a straggler-resilient method that begins training with faster nodes and gradually incorporates slower ones as the model converges, leveraging the interplay between system capabilities and statistical accuracy.

Statistical heterogeneity, where client data is Non-Independent and Identically Distributed (non-IID), is an equally critical challenge. Non-IID data can cause local models to diverge, a phenomenon known as "client drift", which degrades the performance of the global model. Standard FedAvg struggles in these conditions. To improve convergence on heterogeneous data, Reddi *et al*. [2021] proposed adaptive federated optimization methods, including federated versions of ADAGRAD, YOGI, and ADAM, which adapt the learning rate on the server side to stabilize and accelerate training.

## 2.3 Privacy-Enhancing Technologies in FL

While the aforementioned studies have advanced the field of privacy-preserving healthcare, significant gaps remain in handling severe class imbalance and statistical heterogeneity si-multaneously. For instance, the FedHybrid framework Dubey *et al*. [2025] successfully integrates DP for heart disease prediction, but primarily addresses scenarios with balanced or IID data distributions. Our work diverges by explicitly targeting the failure modes of standard approaches on highly imbalanced (approximately 5% positive class) and non-identically distributed (non-IID) clinical data, demonstrating that privacy mechanisms must be coupled with rigorous client-side resampling (SMOTETomek) to avoid model collapse (zero Recall).

Furthermore, distinct from Naresh and Varma [2025], which employs Homomorphic Encryption (HE) to secure model updates, our approach leverages DP. While HE offers the advantage of exact aggregation without noise, it incurs substantial computational and communication overheads that can be prohibitive for resource-constrained edge devices in a federated network. By opting for DP, we offer a lightweight and scalable alternative. Crucially, we advance the state-of-the-art by not merely applying DP, but by analytically quantifying the specific *privacy-utility frontier* (AUC $\approx 0.82$ at $\epsilon \approx 9.0$) required to make this lightweight approach clinically viable for stroke screening.

The landscape of PETs is diverse, and each comes with trade-offs. The surveys by Madathil *et al*. [2025] and Yurdem *et al*. [2024] provide excellent overviews of these technologies, including DP, HE, Secure Multi-Party Computation (SMC), and Trusted Execution Environments (TEEs). A central theme across these studies is the inherent trade-off between privacy and utility: stronger privacy guarantees often come at the cost of reduced model accuracy or increased computational overhead Silva [2025]. For example, while primarily focused on system efficiency, the FedPA framework by Liu *et al*. [2021] acknowledges the need for privacy by testing its robustness to noise added for privacy concerns, implicitly highlighting the privacy-utility trade-off. Consequently, our work is founded on the established challenge of optimizing privacy and utility, seeking not only to acknowledge but also to systematically map their relationship.

# 3 Background and Problem Formulation

Federated Learning enables $N$ clients to collaboratively train a global model while keeping their datasets $D_i$ decentralized. We adopt the standard Horizontal Federated Learning (HFL) setting, where clients share the same feature space but possess distinct sample sets Liu *et al*. [2020]. The global optimization objective is to minimize the weighted average of local loss functions $F_i(w)$, as formally defined in the foundational work of McMahan *et al*. [2017].

While standard FL (via FedAvg) effectively aggregates updates in ideal conditions, it faces two critical challenges in clinical deployments:

- **Statistical Heterogeneity (Non-IID):** Clinical data are rarely identically distributed across institutions. This heterogeneity causes the local objectives $F_i(w)$ to diverge from the global objective, leading to "client drift" and poor convergence. To mitigate this, we employ the **FedProx** algorithm Li *et al*. [2020]. FedProx modifies the local objective by adding a proximal term $\frac{\mu}{2}||w - w^t||^2$, which penalizes local updates that deviate excessively from the global model $w^t$. This regularization is crucial for maintaining stability when training on heterogeneous patient distributions.
- **Privacy Leakage:** Although raw data remains local, gradient updates can inadvertently reveal sensitive patient attributes through reconstruction or membership inference attacks Nguyen *et al*. [2025]. To provide formal guarantees against such leakage, our pipeline integrates DP, specifically the Differentially Private Stochastic Gradient Descent (DP-SGD) algorithm. As detailed in our previous work Tertulino and Alencar [2026], this approach effectively clips and adds noise to gradients before they leave the client's secure environment.

# 4 Methodology

The methodology for this study was designed to predict stroke risk, which is a primary event of CVD, a broad category encompassing various conditions that affect the heart and blood vessels Saxena *et al*. [2016]. The predictive attributes used, such as hypertension, cholesterol levels, and diabetes, are well-established risk factors for the full spectrum of CVDs, not exclusively for stroke Lu [2025]. Therefore, while the specific endpoint of our model is stroke, the problem is correctly framed within the broader, more common context of cardiovascular risk prediction, a task for which **Feed-Forward Neural Networks (FFNNs)** have proven effective Naresh and Varma [2025]. Consequently, the approach aligns with standard medical and machine learning literature practice, where these risk factors indicate overall cardiovascular health Ali *et al*. [2023].

To conduct our analysis, we utilize the publicly available Stroke Prediction Dataset compiled by Fedesoriano [2021]. A summary of the dataset's characteristics is provided in **Table 2**. The dataset is known to have a significant class imbalance, with only a small fraction of instances corresponding to positive stroke cases (approximately 5%). Thereby, the characteristic presents a common and critical challenge in medical data analysis, making it an ideal candidate for evaluating the robustness of our framework.

## 4.1 Overview of the Proposed Pipeline

Our experimental methodology follows a multi-stage pipeline designed to address the key challenges of class imbalance and statistical heterogeneity while preserving privacy. As illustrated in **Figure 1**, the workflow enforces a strict separation of data. For each client, the raw, imbalanced data is first processed locally using SMOTETomek. Subsequently, the local model is trained using the FedProx objective function to handle statistical heterogeneity. Finally, the resulting gradients are privatized using DP-SGD (clipping and noise injection) before the updates are securely transmitted to the central server for aggregation.

**Table 1.** Comparative Analysis of Related Works.

| Reference | Primary Contribution | Key Challenge Addressed | Privacy Technique Used | Domain/Application |
|---|---|---|---|---|
| **Federated Learning in Healthcare** | | | | |
| Zhang *et al*. [2024] | FedCVD benchmark | Lack of real-world benchmarks | N/A (focus on benchmarking) | Cardiovascular Disease |
| Gaber *et al*. [2024] | FedCVD FL Model | Privacy, Data Imbalance | N/A (Privacy by design) | Cardiovascular Disease |
| Ahmed *et al*. [2025] | Scoring-based client selection | Client heterogeneity, Fairness | N/A (focus on selection) | Diabetes Diagnosis |
| Hudaib *et al*. [2025] | Survey of deep learning in FLHC | Comprehensive overview | Discusses privacy as a challenge | General Healthcare, Medical Imaging |
| **Key Challenges in Federated Learning** | | | | |
| Nishio and Yonetani [2019] | FedCS protocol | System Heterogeneity (Stragglers) | N/A (focus on efficiency) | Mobile Edge Computing |
| Reisizadeh *et al*. [2022] | FLANP protocol | System Heterogeneity (Stragglers) | N/A (focus on efficiency) | General FL |
| Reddi *et al*. [2021] | Adaptive optimizers (FedAdam, etc.) | Statistical Heterogeneity (non-IID) | N/A (focus on optimization) | General FL |
| **Privacy-Enhancing Technologies in FL** | | | | |
| Dubey *et al*. [2025] | FedHybrid framework | Privacy, non-IID data | Differential Privacy (Laplace) | Heart Disease Prediction |
| Naresh and Varma [2025] | FL+HE framework | Privacy, Security | Homomorphic Encryption (CKKS) | Heart Stroke Detection |
| Madathil *et al*. [2025] | (Survey) Comparison of PETs | Privacy-Utility Trade-off | Discusses DP, HE, SMC, TEE | General Healthcare |
| Yurdem *et al*. [2024] | (Survey) Overview of privacy | Privacy as a core challenge | Discusses various PETs | General / Cross-Domain |
| Liu *et al*. [2021] | FedPA framework | System Heterogeneity (Stragglers) | Noise for privacy (mentioned) | General FL |
| **Proposed Contribution** | | | | |
| **This Work** | **Robust Privacy Pipeline** | **Dual Heterogeneity (Imbalance + Non-IID)** | **Client-Side SMOTE-Tomek + FedProx + DP-SGD** | **Stroke Screening** |

## 4.2 Experimental Hyperparameters

To ensure a fair and reproducible analysis, the FL process was configured with a set of key hyperparameters held constant across all experimental runs. The chosen values, detailed in **Table 3**, were selected based on common practices in the federated learning literature and preliminary testing to ensure stable model convergence.

The configuration is divided into three main categories. First, the **Federated Learning Configuration** defines the overall structure of the experiment, including the number of participating clients, the total number of training rounds, and the local epochs each client performs per round. Second, the **Model Training Configuration** specifies standard deep learning parameters, such as the optimizer and learning rate. Each client employed the Adam optimizer for local optimization due to its efficiency and adaptivity in deep learning tasks Ismail *et al*. [2025]. Finally, the **FedProx Configuration** details the values tested for the proximal hyperparameter ($\mu$), which was tuned to optimize the algorithm's performance on our non-IID data.

## 4.3 Experimental Framework, Model, and Data

The experimental framework for our study was orchestrated by **Flower**, a flexible, scalable, open-source framework designed for FL research Naseri *et al*. [2024]. Flower's framework-agnostic nature facilitated seamless integration with **PyTorch**, which was used to define and train the machine learning model.

The predictive model for the study is an **FFNN**, an architecture well-suited for capturing non-linear patterns in clinical data and proven effective for stroke prediction tasks Naresh and Varma [2025]. The network architecture, detailed in **Table 4**, consists of an input layer with 15 preprocessed features, two hidden layers with ReLU activations, and a final sigmoid output layer for binary classification.

The **Stroke Prediction Dataset** Fedesoriano [2021] was selected as the experimental testbed as it encapsulates two crit-ical real-world challenges: significant **class imbalance** and **inherent feature diversity**. Our reliance on this standardized public benchmark is a deliberate methodological choice to ensure reproducibility and facilitate direct comparisons with state-of-the-art baselines. While we acknowledge that raw logs from multi-center hospital environments may typically exhibit unstructured noise beyond the scope of this dataset, using this established benchmark allows us to rigorously isolate and validate the privacy-utility trade-off, free from the confounding variables inherent in uncurated data.

**Table 2.** Characteristics of the Stroke Prediction Dataset.

| **General Properties** | **Description** | |
|---|---|---|
| Dataset Name | Stroke Prediction Dataset | |
| Total Instances | 5,110 | |
| Class Distribution | Highly imbalanced ($\approx 5\%$ positive cases for the 'stroke' class) | |
| **Features (10) and Target** | **Type** | **Description** |
| `gender` | Categorical | "Male", "Female", or "Other" |
| `age` | Numeric | Age of the patient |
| `hypertension` | Binary | 0: No hypertension, 1: Has hypertension |
| `heart_disease` | Binary | 0: No heart disease, 1: Has heart disease |
| `ever_married` | Categorical | "Yes" or "No" |
| `work_type` | Categorical | "Private", "Self-employed", "Govt_job", etc. |
| `Residence_type` | Categorical | "Urban" or "Rural" |
| `avg_glucose_level` | Numeric | Average glucose level in blood |
| `bmi` | Numeric | Body Mass Index |
| `smoking_status` | Categorical | "formerly smoked", "never smoked", "smokes", or "Unknown" |
| `stroke` (Target) | Binary | 1: Patient had a stroke, 0: Patient did not have a stroke |

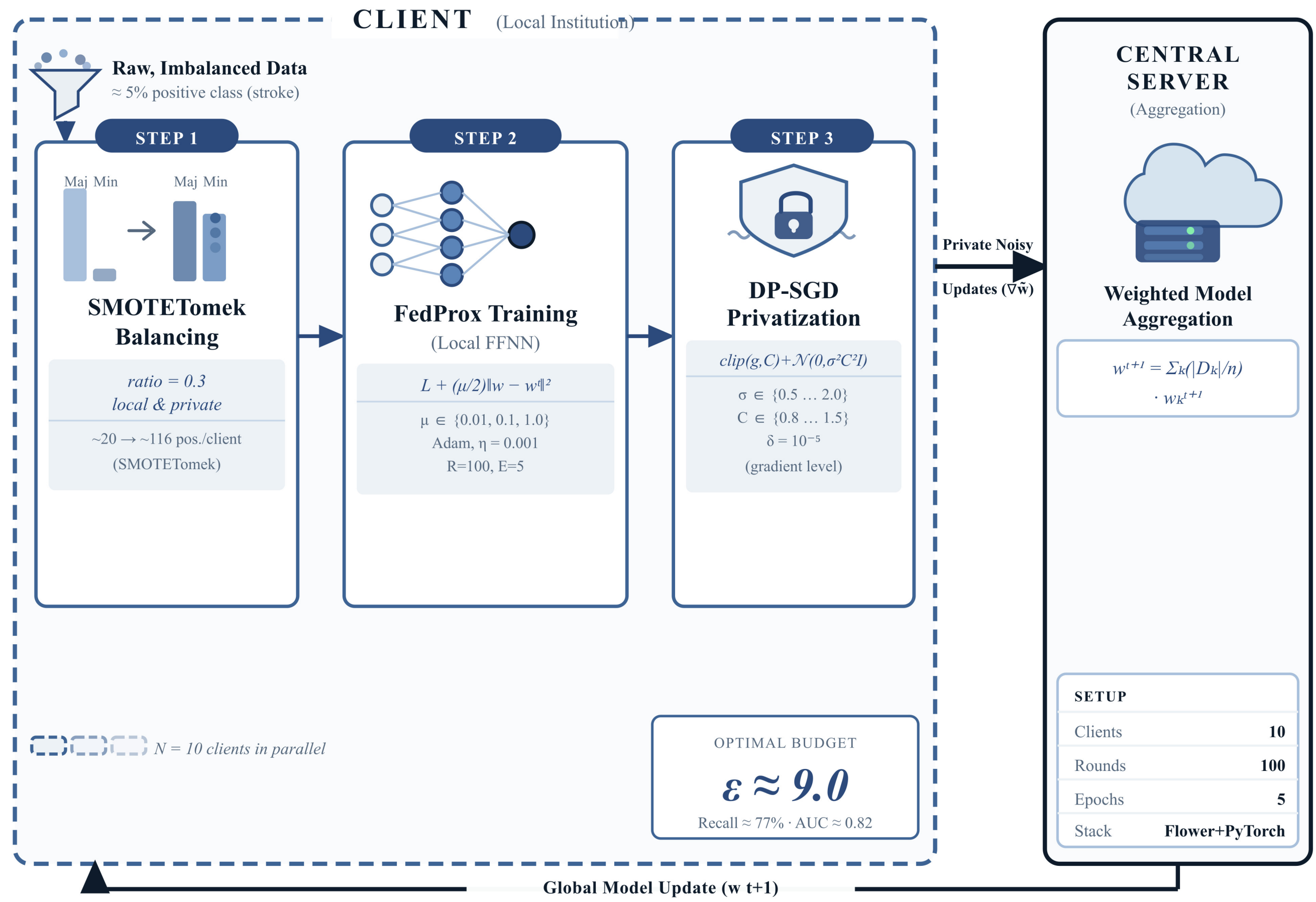

**Figure 1.** Graphical overview of the proposed multi-stage federated learning pipeline. The workflow illustrates the client-side processing steps: (1) **Data Balancing** using SMOTETomek to address class imbalance; (2) **Robust Training** utilizing the FedProx objective function, which includes a proximal term ($\frac{\mu}{2}||w - w^t||^2$) to mitigate statistical heterogeneity; and (3) **Privatization** via DP-SGD, where gradients are clipped and perturbed with Gaussian noise ($\sigma$) before secure transmission. The central server then aggregates these private updates to refine the global model.

**Table 3.** Key Hyperparameters for the Federated Learning Experiments.

| Parameter | Value |
|---|---|
| **Federated Learning Configuration** | |
| Number of Clients ($N$) | 10 |
| Communication Rounds ($R$) | 100 |
| Local Epochs ($E$) | 5 |
| Client Fraction per Round | 1.0 (100%) |
| **Model Training Configuration** | |
| Optimizer | Adam |
| Learning Rate ($\eta$) | 0.001 |
| Batch Size | 32 |
| **FedProx Configuration** | |
| Proximal Mu ($\mu$) | Tuned {1.0, 0.1, 0.01} |
| **Differential Privacy Configuration** | |
| Noise Multiplier ($\sigma$) | Tuned {0.5, 1.0, 1.5, 2.0} |
| Clipping Norm ($C$) | Tuned {0.8, 1.0, 1.2, 1.5} |
| Target Delta ($\delta$) | $10^{-5}$ |

**Table 4.** Feed-Forward Neural Network (FFNN) Model Architecture Summary. The network configuration reflects the processing of 15 input features.

| Layer (Type) | Output Shape | Activation | Params |
|---|---|---|---|
| Input Layer | (None, 15) | - | 0 |
| Hidden Layer 1 (Dense) | (None, 64) | ReLU | 1,024 |
| Hidden Layer 2 (Dense) | (None, 32) | ReLU | 2,080 |
| Output Layer (Dense) | (None, 1) | Sigmoid | 33 |
| **Total Trainable Parameters** | | | **3,137** |

## 4.4 Iterative Refinement of the Federated Learning Pipeline

Our investigation followed an iterative experimental process to develop a robust pipeline. The impact of each methodological refinement is summarized in **Table 5**, which compares the model's performance at a representative privacy configuration across the three key stages of our research.

The initial experiment (**Stage 1**) applied a standard FedAvg algorithm to the raw, imbalanced data. As shown in the table, this approach yielded a misleadingly high accuracy of 87.9% but failed to meet its clinical objective, resulting in a **Recall of 0.0%**. The critical finding from the stage was that standard frameworks are not inherently robust to the severe class imbalance present in the dataset.

To address the aforementioned failure, we integrated the **Synthetic Minority Over-sampling Technique with Tomek Links (SMOTETomek)** Khleel and Nehéz [2023] at the client level (**Stage 2**). The results show a dramatic improvement, with Recall surging to **74.0%**, confirming that the model successfully identified the minority class. Therefore, the data-balancing step was crucial to establishing a clinically viable model.

**Table 5.** Comparative Results of the Experimental Stages. The table highlights the impact of each methodological refinement on model performance at a fixed privacy configuration ($\sigma = 1.0, C = 1.0$).

| Metric | Stage 1 (Initial: FedAvg) | Stage 2 (+ SMOTE-Tomek) | Stage 3 (+ FedProx) |
|---|---|---|---|
| Accuracy | 87.9% | 72.8% | 72.8% |
| **Recall** | **0.0%** | **74.0%** | **77.0%** |

In addition, to optimize for the non-IID nature of the data, we replaced FedAvg with the FedProx algorithm (**Stage 3**). The final refinement further improved the key clinical metric, with the Recall increasing to **77.0%**, demonstrating FedProx's superior ability to handle statistical heterogeneity and lead to a more effective final model.

## 4.5 Federated Data Partitioning and Local Balancing

To rigorously simulate an HFL scenario, we adopted a strict pipeline that respects client boundaries during data processing. The experimental setup followed this specific order of operations:

1. **Global Stratified Split:** The raw dataset was first split into training (80%) and testing (20%) sets. The testing set was isolated at the server for global evaluation and was not exposed to the clients or the balancing algorithms.
2. **Partitioning and Statistical Heterogeneity:** The training set was distributed among $N = 10$ clients using a random horizontal partitioning scheme. While random partitioning is technically IID with respect to the feature distribution, the extreme sparsity of the positive class ($\approx$5%) means each client's partition holds, on average, only $\approx$20 positive cases versus $\approx$388 negative cases, resulting in widely varying local class prevalences. This label-sparsity-driven instability introduces statistical heterogeneity across clients that, in practice, challenges the aggregation algorithm in a manner analogous to non-IID conditions.
3. **Local Client-Side Balancing:** Crucially, the **SMOTE-Tomek** hybrid sampling was applied *locally* and *independently* by each client on their private partition $D_i$. This ensures that synthetic samples are generated solely from local distributions, preventing data leakage between institutions. We used a sampling ratio of $0.3$ (minority class resampling to 30% of the majority), empirically determined to offer the best trade-off between Precision and Recall.

On average, each client holds approximately $408$ training samples. Given the global prevalence, this results in an average of approximately $\approx 20$ positive cases (Stroke) against $\approx 388$ negative cases (Healthy) per client. This scarcity creates a scenario in which local models are highly susceptible to overfitting and divergence, mirroring the challenges of Non-IID clinical data. **Table 6** details the data distribution before and after the local balancing process.

**Table 6.** Average Data Distribution per Client ($N = 10$). Balancing (SMOTE Ratio 0.3) is applied locally to ensure privacy.

| State | Negative (Mean) | Positive (Mean) | Total Samples (Mean) |
|---|---|---|---|
| Raw Partition | 388 | $\approx 20$ | $\approx 408$ |
| After Local SMOTE | 388 | $\approx 116$ | $\approx 504$ |

This decentralized balancing approach directly impacts the DP accountant, as the privacy budget is consumed based on the size of the local datasets that the optimizer actually sees during training.

**Privacy unit and local oversampling.** A clarification is warranted for DP readers. Our DP-SGD implementation provides *example-level* privacy: the budget $\epsilon$ bounds what can be inferred about any single training record processed by the local optimizer. Because SMOTETomek is applied entirely within each client's private environment prior to training, no synthetic sample leaves the client's boundary. However, since synthetic records are derived from real patient data, they should not be treated as independent individuals for privacy accounting. Accordingly, the privacy budget is computed with respect to the post-balancing local dataset size, a standard and conservative approximation consistent with established DP-SGD practice. Readers primarily interested in *client-level* DP guarantees should note that such analysis requires a separate, amplification-based framework, which is left for future work.

## 4.6 Data Preprocessing

Before model training, a series of preprocessing steps was performed to ensure data quality and compatibility with the neural network architecture. These tasks are crucial for addressing inconsistencies in the raw data and for enhancing the performance of the machine learning model.

The preprocessing pipeline involved the following key tasks:

- **Handling Missing Values:** The BMI (Body Mass Index) feature contained many missing entries. Thus, we employed a mean imputation strategy to handle missing values in the BMI feature, replacing each missing value with the column mean. Hence, the mean imputation strategy preserves the overall distribution of the feature without discarding valuable data from affected instances.
- **Encoding Categorical Features:** The dataset included several categorical features, such as gender, work_type, and smoking_status. These textual features were converted to numerical values using one-hot encoding. The process creates new binary columns for each category, allowing the model to interpret the information without implying an ordinal relationship between categories.
- **Feature Scaling (Normalization):** The numerical features in the dataset, such as age, avg_glucose_level, and BMI, operate on vastly different scales, which can lead to slower convergence and cause the model to assign undue importance to features with larger numerical ranges. To mitigate these issues, we applied Standard Scaling (also known as Z-score normalization) to all numerical features Wang *et al*. [2024]. Thus, the process rescales each feature to have a mean of 0 and a standard deviation of 1, ensuring that all features contribute equally to the model's learning process. Crucially, the scaling was fitted only on the training data to prevent data leakage from the test set.
- **Addressing Class Imbalance with SMOTETomek:** Our initial experiments revealed that the severe class imbalance (approx. 5% positive cases) prevented the model from learning to identify the minority class, resulting in a Recall of zero. To address the critical challenge of class imbalance, we chose the SMOTETomek. While the standard SMOTE algorithm effectively increases Recall by oversampling the minority class, it often achieves significantly lower precision, as the samples can create noisy or overlapping class boundaries.

The resulting cleaned and balanced data partitions formed the basis for the federated training experiments described in the following sections.

## 4.7 Differential Privacy Implementation

We implement client-side DP using the **Opacus** library, which integrates seamlessly with **PyTorch**. The mechanism is based on DP-SGD, which involves two key steps applied by each client during local training:

1. **Per-sample Gradient Clipping:** Before averaging, the L2 norm of each gradient is clipped to a maximum value, C. The clipping process bounds the maximum influence a single data point can have on the update, with the clipping threshold controlled by the "max_grad_norm" parameter.

$$\tilde{g}_t(x_i) = \frac{g_t(x_i)}{\max\left(1, \frac{\|g_t(x_i)\|_2}{C}\right)}$$

2. **Noise Addition:** Gaussian noise with a standard deviation proportional to the clipping bound C and a noise_multiplier sigma is added to the clipped, averaged gradients for a batch B.

$$\bar{g}_t = \frac{1}{|B|} \sum_{i \in B} \tilde{g}_t(x_i) + \mathcal{N}(0, \sigma^2 C^2 \mathbf{I})$$

The server then aggregates these noisy gradients. The level of privacy is determined by the combination of these parameters and is quantified by the privacy budget ($\epsilon$).

## 4.8 Privacy Accounting and Hyperparameters

To ensure reproducibility of the privacy budget ($\epsilon$), we specify the exact accounting parameters derived from the experimental setup. The privacy cost was tracked using the **Rényi Differential Privacy (RDP) accountant** provided by the Opacus library.

The cumulative noise was calculated as the sum of Gaussian mechanisms across the total number of training steps. The specific parameters for the privacy accountant are:

- **Sampling Rate ($q$):** Calculated locally for each client as $q = \frac{B}{|D'_i|}$, where $B = 32$ is the batch size and $|D'_i| \approx 504$ is the size of the local dataset after SMOTE augmentation (Ratio 0.3). Thus, the sampling rate is $q \approx 0.063$.
- **Steps per Round:** Each client performs $E = 5$ local epochs per round. With $\approx 15$ steps per epoch (based on dataset size), this results in $\approx 75$ optimization steps per communication round.
- **Total Steps ($T$):** Over $R = 100$ communication rounds, the total number of DP-SGD steps contributing to the privacy loss is $T \approx 7,500$.
- **Delta ($\delta$):** We adopted a standard value of $\delta = 10^{-5}$, ensuring $\delta \ll \frac{1}{|D|}$, to bound the probability of privacy breach.

Given these parameters and the noise multiplier $\sigma$, the final $\epsilon$ is derived by minimizing the RDP budget over orders $\alpha$:

$$\epsilon(\delta) = \min_{\alpha} \left( \frac{\mathcal{L}_{RDP}(\alpha)}{\alpha - 1} + \frac{\log(1/\delta)}{\alpha - 1} \right) \quad (1)$$

where $\mathcal{L}_{RDP}(\alpha)$ is the cumulative Rényi divergence for the composed mechanism.

### 4.9 Experimental Design

Our core experiment is a parameter sweep designed to map the performance surface as a function of DP settings. The full experimental procedure is outlined in **Algorithm 1**.

- **Independent Variables:**
  We define a grid of values for noise_multiplier and max_grad_norm to systematically explore the parameter space.
- **Metrics:** For each configuration, we record the final global model's Accuracy, Recall, and the Area Under the Receiver Operating Characteristic Curve (ROC-AUC) on a held-out test set. We emphasize Recall for clinical sensitivity and ROC-AUC to assess the model's discriminative ability independent of the decision threshold.

While **Algorithm 1** outlines the high-level experimental loop for the parameter sweep, the core distinction of the FedProx method lies within the client-side training procedure. To provide a clear and reproducible account of our methodology, we detail the local update process in **Algorithm 2**.

Therefore, the Client-Side Training algorithm describes the steps each client performs in each communication round. It explicitly shows how the standard loss function is augmented with the FedProx proximal term (Line 6), which penalizes deviations from the global model's weights ($w^t$). Thus, the modification is the key mechanism for mitigating client drift in non-IID settings. The algorithm then details the subsequent gradient computation and model update steps, which are performed by the differentially private (DP-SGD) optimizer, thus combining the robustness of FedProx with the privacy guarantees of DP.

**Algorithm 1** Granular Analysis with Optimized FedProx

**Require:** Client datasets $\{D_1, \dots, D_N\}$, learning rate $\eta$, local epochs $E$, communication rounds $R$.
**Require:** Set of proximal hyperparameters $\mathcal{S}_\mu$.
**Require:** Set of noise multipliers $\mathcal{S}_\sigma$, Set of clipping norms $\mathcal{S}_C$.
**Ensure:** A log of performance metrics, **MetricsLog**.

1: Initialize an empty list **MetricsLog**.
2: **for all** proximal hyperparameter $\mu \in \mathcal{S}_\mu$ **do**
3:     **for all** noise multiplier $\sigma \in \mathcal{S}_\sigma$ **do**
4:         **for all** clipping norm $C \in \mathcal{S}_C$ **do**
5:             Initialize global model $w_0$.
6:             **for** each communication round $t = 0, \dots, R-1$ **do**
7:                 Select a subset of clients $S_t$.
8:                 $n_t \leftarrow \sum_{k \in S_t} |D_k|$ ▷ Total samples in round
9:                 **for all** client $k \in S_t$ **in parallel do**
10:                     $w_k^{t+1} \leftarrow$ ClientUpdateFedProx$(w_t, D_k, E, \eta, \sigma, C, \mu)$ ▷ See Alg. 2
11:                 **end for**
12:                 $w_{t+1} \leftarrow \sum_{k \in S_t} \frac{|D_k|}{n_t} w_k^{t+1}$ ▷ Weighted Aggregation
13:             **end for**
14:             Evaluate final model $w_R$ to get metrics $M$.
15:             Calculate privacy budget $\epsilon$ for $(\sigma, C, R)$.
16:             Append $((\mu, \sigma, C), (M, \epsilon))$ to **MetricsLog**.
17:         **end for**
18:     **end for**
19: **end for**
20: **return MetricsLog**

**Algorithm 2** Client-Side Training with Proximal Term

**Require:** Global model weights $w^t$, local data $D_k$, epochs $E$, learning rate $\eta$, DP parameters $(\sigma, C)$, proximal term $\mu$.
**Ensure:** Updated local model weights $w_k^{t+1}$.

1: Initialize local model $w \leftarrow w^t$.
2: Attach DP-SGD engine to model with parameters $(\sigma, C)$.
3: **for** each local epoch $e = 1, \dots, E$ **do**
4:     **for** each batch $b \in D_k$ **do**
5:         Calculate standard loss $L_{std} \leftarrow \text{Loss}(w, b)$.
6:         Calculate proximal term $L_{prox} \leftarrow \frac{\mu}{2} \|w - w^t\|^2$.
7:         Calculate total loss $L_{total} \leftarrow L_{std} + L_{prox}$.
8:         Compute gradient $\nabla L_{total}$.
9:         Update local model $w \leftarrow w - \eta \nabla L_{total}$ via DP-SGD optimizer.
10:     **end for**
11: **end for**
12: **return** $w_k^{t+1} \leftarrow w$

## 5 Results and Discussion

Our investigation into the privacy-utility trade-off was conducted through a multi-stage experimental process, culminating in a rigorous statistical validation. Initial experiments revealed the critical impact of class imbalance, in which standard approaches yielded zero Recall, guiding methodological refinements that ultimately enabled a robust comparative analysis of FL algorithms under DP.

## 5.1 Statistical Robustness and Performance Analysis

To address the stochastic nature of DP and ensure the reliability of our findings, we conducted a robustness analysis based on five independent experimental runs with varying random seeds. **Table 7** summarizes the aggregated performance metrics at the identified optimal privacy budget ($\epsilon \approx 9.0$).

**Table 7.** Robustness Analysis of the Proposed Pipeline (SMOTETomek-FedProx) under Differential Privacy ($\epsilon \approx 9.0$). Results are aggregated from 5 independent experimental runs to account for DP noise stochasticity.

| Metric | Mean | Std. Dev. | 95% Conf. Interval |
|---|---|---|---|
| ROC-AUC | 0.8219 | ± 0.0216 | [0.7951, 0.8487] |
| Recall | 0.7700 | ± 0.1068 | [0.6374, 0.9026] |
| Precision | 0.7466 | ± 0.0442 | [0.6917, 0.8015] |
| F1-Score | 0.7581 | ± 0.0603 | [0.6832, 0.8330] |
| Accuracy | 0.7280 | ± 0.0173 | [0.7065, 0.7495] |

The results demonstrate that the proposed pipeline achieves high efficacy. Crucially, the model maintains a **Mean F1-Score of 0.7581**, indicating a successful balance between sensitivity (Recall $\approx 0.77$) and precision ($\approx 0.75$). This finding refutes the common assumption that privacy-preserving models on imbalanced data must necessarily sacrifice precision to achieve safety Suriyakumar *et al*. [2021]. The standard deviations remain within acceptable limits, confirming that the FedProx algorithm effectively stabilizes the global model despite noise introduced by DP-SGD.

## 5.2 Discriminative Capability: ROC-AUC Analysis

While point metrics, such as the F1-score, depend on the decision threshold, the Area Under the ROC Curve (ROC-AUC) provides a threshold-independent measure of the model's discriminative power.

As illustrated in **Figure 2**, the model exhibits a stable performance with an AUC of approximately $0.82$ across the privacy spectrum. This stability indicates that the combination of local SMOTETomek balancing and FedProx regularization enables the global model to learn robust feature representations for stroke risk, which are resilient to the perturbations required for DP.

## 5.3 Operational Trade-off and Privacy Budget Analysis

Based on the granular analysis of the privacy-utility frontier, we define the *optimal operational region* for this clinical application as the privacy budget interval $\epsilon \in [8.0, 10.0]$.

Within this range, the model achieves a Pareto-efficient balance, maintaining high clinical utility (Recall $> 75\%$, AUC $\approx 0.82$) at a privacy budget centered on a single-digit value ($\epsilon \approx 9.0$). Below this region ($\epsilon < 5.0$), clinical utility degrades sharply, rendering the model unsafe for screening purposes due to the increased risk of false negatives.

**Practical Implications of $\epsilon \approx 9.0$:** While theoretical DP literature often explores $\epsilon < 1.0$, real-world healthcare applications dealing with high-dimensional, heterogeneous data often require slightly more relaxed budgets to be clinically viable. A budget of $\epsilon \approx 9.0$ offers a pragmatic trade-off: it provides mathematical protection against reconstruction and membership inference attacks, significantly superior to non-private training ($\epsilon = \infty$), while preserving the high sensitivity required for effective disease screening.

## 5.4 Comparative Performance: FedProx vs. FedAvg

To validate the choice of aggregation algorithm in our non-IID setting, we compared the optimized FedProx against the standard FedAvg.

The comparative analysis in **Figure 3** reveals that while standard FedAvg performs competitively, achieving Recall scores surpassing $75\%$ in moderate privacy regimes, **FedProx demonstrates superior stability**. The proximal term in FedProx effectively regularizes local updates, reducing variance caused by skewed data distributions (client drift). Consequently, FedProx maintains a consistent performance advantage, offering a more reliable aggregation mechanism for safety-critical medical applications where stability under noise is as crucial as raw performance.

# 6 Conclusion and Future Directions

## 6.1 Conclusion

The present research provides a comprehensive analysis of the privacy-utility trade-off in FL applied to imbalanced clinical data for cardiovascular risk prediction. Our iterative experimental process demonstrated that standard privacy-preserving FL approaches fail in realistic healthcare settings characterized by severe class imbalance, resulting in models with negligible clinical utility (zero Recall).

To overcome this, we developed and validated a robust multi-stage pipeline that integrates local SMOTETomek balancing and the FedProx aggregation algorithm. Our robust statistical analysis, aggregated over five independent experimental runs, confirms the effectiveness of this approach. The proposed model achieves a **Mean ROC-AUC of** $0.82$ and a **Mean F1-Score of** $0.76$ at a privacy budget of $\epsilon \approx 9.0$. These results represent a significant advancement, demonstrating that it is possible to maintain high clinical sensitivity (Recall $\approx 77\%$) and precision ($\approx 75\%$) simultaneously, refuting the assumption that privacy noise inevitably renders models imprecise on imbalanced data.

Ultimately, this work serves as a practical blueprint for institutions aiming to develop trustworthy collaborative AI. It highlights that Privacy-Enhancing Technologies (PETs) cannot operate in isolation; they must be integrated into a data-centric pipeline that actively addresses local data sparsity and statistical heterogeneity to ensure both security and clinical efficacy.

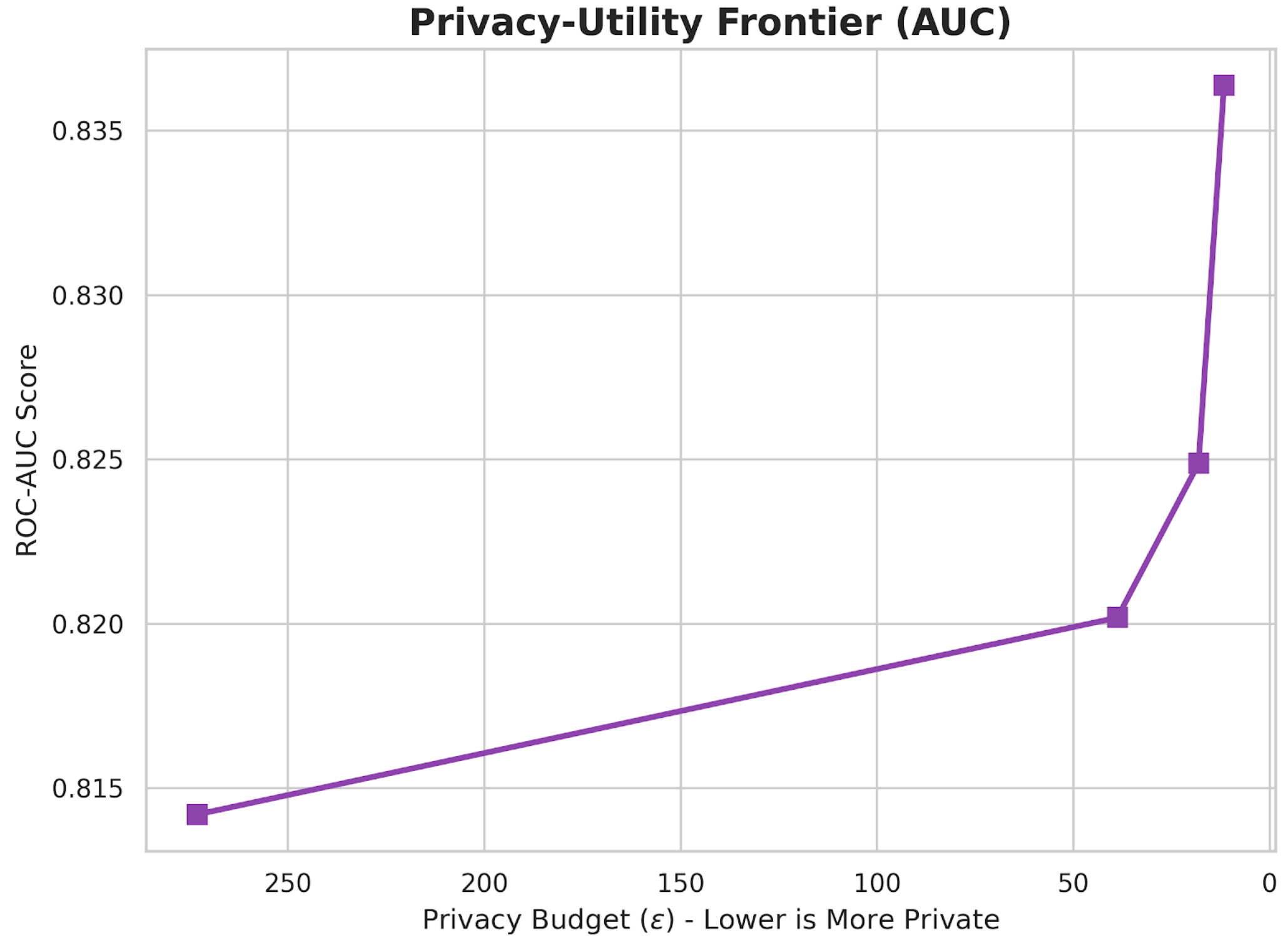

**Figure 2.** Robust Privacy-Utility Frontier measured by ROC-AUC. The plot demonstrates that the optimized FedProx model maintains a strong discriminative capability (Mean AUC $\approx 0.82$) across varying privacy budgets, showing resilience to the noise introduced by Differential Privacy.

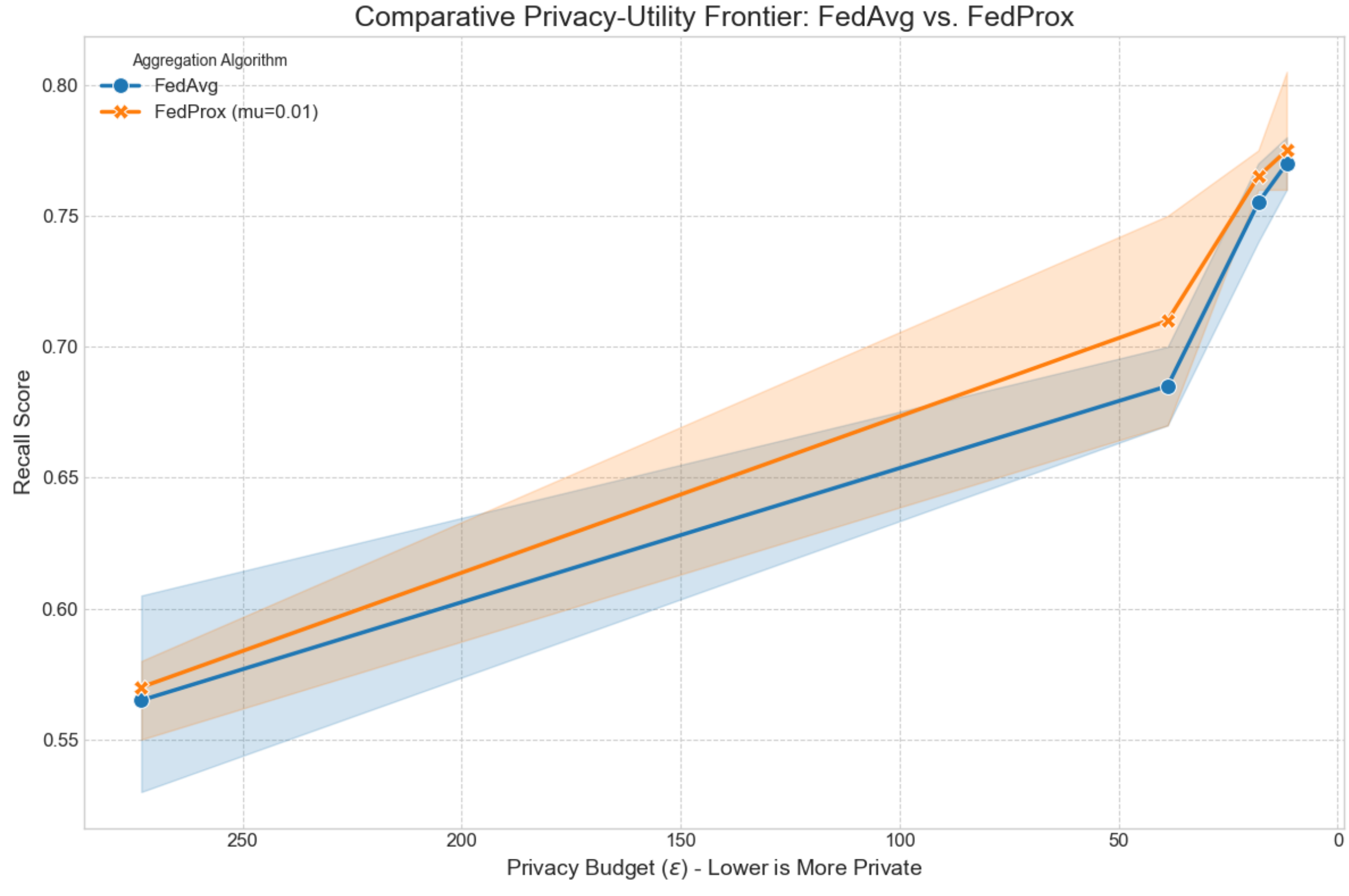

**Figure 3.** Comparative Performance: FedAvg vs. Optimized FedProx ($\mu = 0.01$). While FedAvg demonstrates competitive utility (Recall $> 75\%$), FedProx exhibits superior stability and consistency across the privacy spectrum, particularly in maintaining robust performance under stricter privacy budgets.

## 6.2 Future Work

Our findings open several compelling avenues for future research to further optimize this privacy-utility frontier:

- **Advanced Client Selection:** Future work should evaluate dynamic client selection protocols, such as scoring-based mechanisms Ahmed *et al*. [2025], to balance statistical contribution with system efficiency in heteroge-neous networks.

- **Adaptive Federated Optimization:** While FedProx proved robust, exploring server-side adaptive optimizers such as FedAdam or FedYogi Reddi *et al*. [2021] could accelerate convergence and improve the privacy budget efficiency by reducing the required training rounds.
- **Hybrid Privacy Models:** Extending the analysis to Hybrid PETs is a critical next step. Combining DP with Secure Multi-Party Computation (SMC) or Homomorphic Encryption (HE) Naresh and Varma [2025] could offer a "best of both worlds" solution, minimizing the noise injection required for privacy while managing the computational overhead of cryptographic methods.

# Declarations

## Authors' Contributions

Rodrigo Tertulino is the sole author of the paper.

## Competing interests

The author declares no competing interests.

## Acknowledgements

The author would like to acknowledge the support of the Software Engineering and Automation Research Laboratory (LaPEA), where the research was developed and conducted. The infrastructure and resources provided were crucial for completing the work. In the preparation of this work, the author used Claude (Anthropic) and Grammarly to review text fluency and language, spelling, and grammar errors. These tools were not used to generate content. After using these tools, the author reviewed and edited the content as necessary, taking full responsibility for its accuracy.

## Funding

This research was not funded.

## Availability of data and materials

The dataset used and analyzed during the current study is publicly available. The "Stroke Prediction Dataset", created by Fedesoriano [2021], was sourced from the Kaggle repository. It can be accessed directly via the following URL: `https://www.kaggle.com/datasets/fedesoriano/stroke-prediction-dataset`

10.48550/arxiv.2411.07050.

# A Code and Key Functions

The appendix details the core components of the final Python implementation used to conduct the experiments described in this paper. The primary goal is to provide methodological clarity and ensure the reproducibility of our analysis. Our experimental framework was built upon well-established, open-source libraries, including **Flower** for federated learning orchestration, **PyTorch** for neural network modeling, **Opacus** for DP, and **imbalanced-learn** for data balancing.

The following sections present commented code snippets of the essential functions that govern our final experimental pipeline, including: (1) client-side data partitioning and balancing with the hybrid SMOTETomek technique, (2) the federated client training loop incorporating the FedProx algorithm with DP-SGD, and (3) the function for calculating the privacy budget. The complete, executable scripts are available from the author upon request.

```
# --- 1. Data Partitioning with SMOTETomek ---
def create_client_dataloaders(X_train, y_train, num_clients, seed):
    """
    Distributes training data and applies SMOTETomek locally.
    Crucial for Privacy: Balancing happens AFTER partitioning.
    """
    data_len = len(X_train)
    indices = list(range(data_len))
    len_per_client = data_len // num_clients
    client_dataloaders = []

    # Use a moderate ratio (0.3) to prevent a high False Positive rate
    smote = SMOTETomek(sampling_strategy=0.3, random_state=seed)

    for i in range(num_clients):
        # Partitioning (Non-IID simulation via sparsity)
        start = i * len_per_client
        end = (i + 1) * len_per_client
        idx = indices[start:end]
        X_client, y_client = X_train[idx], y_train[idx]

        # Local Balancing (Private Operation)
        # Apply SMOTE only if class distribution allows
        if np.sum(y_client) > 1:
            try:
                X_res, y_res = smote.fit_resample(X_client, y_client)
            except ValueError:
                # Fallback for extremely small partitions
                X_res, y_res = X_client, y_client
        else:
            X_res, y_res = X_client, y_client

        dataset = TensorDataset(
            torch.tensor(X_res, dtype=torch.float32),
            torch.tensor(y_res, dtype=torch.float32).view(-1, 1)
        )
        client_dataloaders.append(DataLoader(dataset, batch_size=32, shuffle=True))

    return client_dataloaders
```

Listing 1: Data Partitioning with Local SMOTETomek Balancing

```
# --- 2. Differentially Private Federated Client with FedProx ---
class DPClient(fl.client.NumPyClient):
    def fit(self, parameters, config):
        self.set_parameters(parameters)

        # Safety check for Opacus compatibility to avoid pickling errors
        if not ModuleValidator.is_valid(self.model):
            self.model = ModuleValidator.fix(self.model)

        privacy_engine = PrivacyEngine()
        optimizer = optim.Adam(self.model.parameters(), lr=0.001)

        # Attach DP-SGD Engine (Clipping + Noise)
        model, optimizer, loader = privacy_engine.make_private(
            module=self.model, optimizer=optimizer,
            data_loader=self.loader,
            noise_multiplier=1.0, max_grad_norm=1.0
        )

        criterion = nn.BCELoss()
        model.train()
        mu = config.get("proximal_mu", 0.01)
        global_params = [torch.tensor(p) for p in parameters]

        # Local Training Loop
        for _ in range(config.get("local_epochs", 5)):
            for inputs, labels in loader:
                optimizer.zero_grad()
                outputs = model(inputs)

                # Standard Loss
                loss = criterion(outputs, labels)

                # FedProx Proximal Term Calculation
                proximal_term = 0.0
                for local_p, global_p in zip(model.parameters(), global_params):
                    proximal_term += torch.sum((local_p - global_p) ** 2)

                # Total Loss (Original + Proximal Penalty)
                loss += (mu / 2) * proximal_term

                loss.backward()
                optimizer.step()

        return self.get_parameters(config={}), len(self.loader.dataset), {}
```

Listing 2: Differentially Private FedProx Client Update

```
# --- 3. Privacy Budget Calculation ---
from opacus.accountants import RDPAccountant

```

```python
def calculate_epsilon(total_steps, sample_rate, noise_multiplier, delta):
    """
    Calculates epsilon using Renyi Differential Privacy (RDP).
    """
    accountant = RDPAccountant()

    # Accumulate privacy cost over all training steps
    for _ in range(total_steps):
        accountant.step(
            noise_multiplier=noise_multiplier,
            sample_rate=sample_rate
        )

    return accountant.get_epsilon(delta)
```

Listing 3: Privacy Budget Calculation (RDP)